\documentclass[
reprint,
unsortedaddress,
showpacs,
nofootinbib,
nobibnotes,
 amsmath,amssymb,
 aps,
pra,
floatfix,
]{revtex4-1}

\usepackage{graphicx}
\usepackage{amsmath}
\usepackage{amssymb}
\usepackage{amsfonts}
\usepackage{latexsym}
\usepackage{setspace}
\graphicspath{{Figures/}}
\newcommand{\Mloc}{\langle M_{\text{loc}}\rangle}
\newcommand{\Mlocparallel}{\langle M_{\text{loc}}^\parallel\rangle}
\newcommand{\Mloclambda}{\langle M_{\text{loc}}(\lambda)\rangle}
\begin{document}
\title{Modified optical absorption of molecules on metallic nanoparticles at sub-monolayer coverage}
\author{Brendan L. Darby}
\author{Baptiste Augui\'e}
\author{Matthias Meyer}
\author{Andres E. Pantoja}
\author{Eric C. Le Ru} \email{Eric.LeRu@vuw.ac.nz}
\affiliation{The MacDiarmid Institute for Advanced Materials and Nanotechnology\\
School of Chemical and Physical Sciences\\ Victoria University of Wellington\\
PO Box 600 Wellington, New Zealand}
\date{\today}

\begin{abstract}
Enhanced optical absorption of molecules in the vicinity of metallic nanostructures is key to a number of surface-enhanced spectroscopies and of great general interest to the fields of plasmonics and nano-optics. Yet, experimental access to this absorbance has long proven
elusive. We here present direct measurements of the intrinsic absorbance of dye-molecules adsorbed onto silver nanospheres, and crucially, at sub-monolayer concentrations where dye--dye interactions become negligible. With a large detuning from the plasmon resonance, distinct shifts and broadening of the molecular resonances reveal the intrinsic properties of the dye in contact with the metal colloid, in contrast to the often studied strong-coupling regime where the optical properties of the dye-molecules cannot be isolated. The observation of these shifts together with the ability to routinely measure them has broad implications in the interpretation of experiments involving resonant molecules on metallic surfaces, such as surface-enhanced spectroscopies and many aspects of molecular plasmonics.
\end{abstract}

\maketitle 


Over the last two decades, the optical properties of metallic nanoparticles (NPs) of various sizes and shapes~\cite{2008MulvaneyCSR} have been exploited in numerous contexts: for sub-wavelength light manipulation, as nano-antennas~\cite{2010SchullerNATMAT} or as building blocks for meta-materials~\cite{2012KauranenNAT}; for ultra-sensitive spectroscopy, such as surface-enhanced fluorescence~\cite{2006KuhnPRL,2007LeRuJPCCFluo} (SEF), and Raman spectroscopy~\cite{2009book,1985MoskovitsRMP,2006Aroca} (SERS); or as sensors~\cite{2008BarhoumiJACS,2011LeruNL,2012BroloNAT}.
More recent applications are emerging, leveraging this understanding of plasmonic optical properties and relying on the rigorous modelling offered by electromagnetic theory: we can cite for example the development of new types of chiroptical spectroscopies~\cite{Gov10}, the enhancement of photochemical reactions on surfaces~\cite{2011XuOE,2013KleinmanPCCP,2014GallowayJPCC,2011LinicNAT}, enhanced light emission in LEDs~\cite{2013Lozanolsa}, or
improved photovoltaics~\cite{2010AtwaterNAT,2014MubeenACSNANO,2014SheldonSci}. Many of these existing and emerging applications are underpinned by the fact that the optical (electronic) absorption of molecules on the surface of metallic NPs is enhanced. But spectral changes induced by molecular adsorption are often ignored because of the experimental challenge of measuring surface absorbance spectra on nanoparticles,
despite early attempts more than 30 years ago \cite{1981CraigheadOL}.

This question is not directly addressed in the great number of recent studies devoted to the topic of strong-coupling between plasmons and molecules~\cite{2004WiederrechtNL,2006HaesJACS,2007ZhaoJACS,2008FofangNL,2010NiJACS,2010NiNL,2010DavisPRB,2011ValmorraAPL,2012ChenJPCC,2013ZenginSREP,2013FangAPL,2013SchlatherNL,2014FaucheauxJPCC,2014AntosiewiczACSPHOT,2014CacciolaACSNANO}; in this regime, the plasmon--molecule interaction is evidenced by a typical anti-crossing of the two resonances as a function of detuning~\cite{2008FofangNL,2013SchlatherNL},
but in such a strongly interacting system the molecular response cannot be isolated.
Moreover, in such studies the dye concentration is often large (typically monolayer coverage and above) to maximize dye/plasmon interactions. Dye--dye interactions cannot therefore be neglected and are expected to induce resonance shifts of the dye layer independently of any plasmonic effects; in fact many studies specifically work with $J$-aggregates rather than isolated dyes~\cite{2004WiederrechtNL,2008FofangNL,2010NiJACS,2010NiNL,2013ZenginSREP,2013SchlatherNL}. As a result, the intrinsic effect (chemical and/or electromagnetic) of the NP on an isolated adsorbed molecule cannot be elucidated.
To this aim, we will show that it is necessary to measure the absorbance of the dye when the dye and plasmon resonances are detuned (to avoid dye--plasmon interaction effects) and at low surface coverage (to avoid dye--dye interaction effects).
This low surface coverage poses a significant experimental challenge because the dye absorbance is then very small and obscured by the large optical response (absorption and scattering) of the NPs.

We here propose to measure NP/dye solutions inside an integrating sphere to obtain the modified absorbance spectra of the commonly used chromophores Rhodamine 700, Rhodamine 6G, Nile Blue and Crystal Violet adsorbed onto spherical Ag NPs in solution, at concentrations of the order of only 10\,nM.
This approach enables the first direct experimental observation of the intrinsic surface absorbance of dyes on NPs in a regime where dye--dye and dye--plasmon resonance interactions are negligible. The observation of noticeable changes in the absorption spectrum of several common dyes clearly demonstrates that spectral shifts are the norm rather than the exception and therefore crucial to any quantitative interpretation of any SERS, SEF, or molecular plasmonics experiments
involving dyes.
%
\section* {Predictions from Electromagnetic Theory}
\begin{figure*}
\centering
\includegraphics[width=15cm]{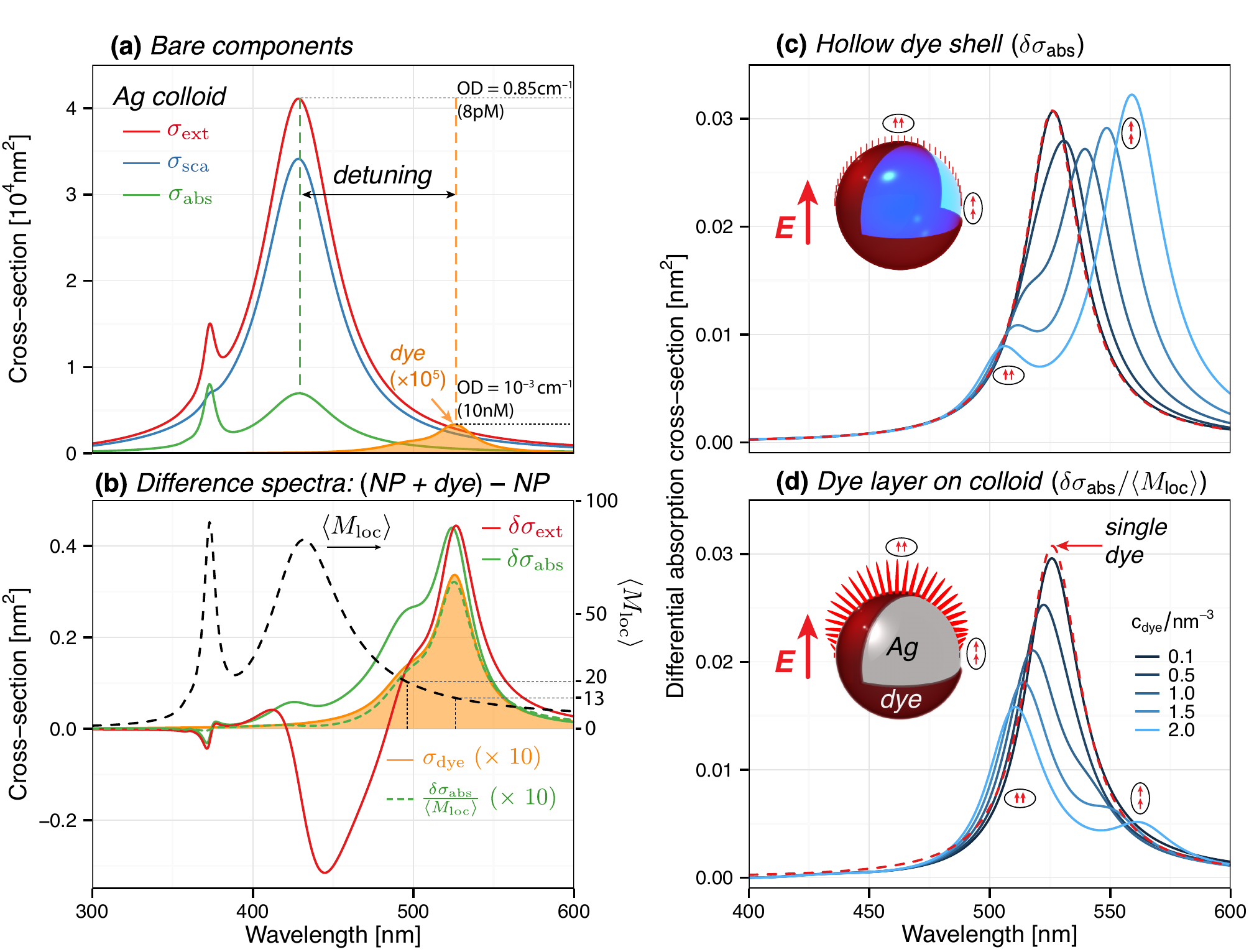}
\caption{
\setstretch{1}
Mie theory simulations for the optical response of dye coated silver spheres. (a) Extinction (red), absorption (green), and scattering (blue) cross-sections for a 30\,nm-radius silver nanosphere immersed in water. For comparison, the absorption cross-section of a typical dye (with a resonance at 526\,nm and a vibronic shoulder)
is shown in orange, with a scaling factor of $10^5$. The peak optical densities (OD) are indicated for concentrations to our experiments: 8\,pM Ag colloids and 10\,nM dye. (b) Difference spectra obtained by subtracting the response of the bare silver sphere shown in (a) from the same colloid coated with a 1\,nm shell of dye (concentration 0.1\,nm$^{-3}$). The dashed black line corresponds to $\Mloc$, the surface-averaged field intensity enhancement factor (right-axis scale).
The corrected differential absorbance $\delta\sigma_\mathrm{abs}(\lambda)/\Mloclambda$ matches closely the reference dye absorbance. (c-d) Differential absorption spectra as a function of dye concentration (from 0.1 to 2.0\,nm$^{-3}$) in the shell.
A single Lorentzian resonance at 526\,nm is used here to highlight peak shifts.
The fictitious hollow shell with the NP replaced
by water (c) is used as a toy model to understand the more relevant case of a dye layer on an Ag sphere (d). The differential absorption spectra in (d) have been normalised by $\Mloc$ to permit a direct comparison with the spectrum of a single dye in water (shown as dashed red lines). The spectra in (c-d) split into two
resonances associated with side-by-side and head-to-tail dipolar interactions, as dye-dye interactions become important. 
The insets show the polarisation ellipse of the local electric field (at 530\,nm). For a metal NP, it is strongly enhanced and mostly radial, hence
favouring side-by-side interactions.}
\label{Fig1}     
\end{figure*} 
We present in Fig.~\ref{Fig1} a general description of the problem, using Mie theory to model the optical response of silver colloids, and a standard isotropic effective medium ~\cite{2004WiederrechtNL,2008FofangNL,2010NiNL,2012ChenJPCC,2013ZenginSREP,2013SchlatherNL,2014AntosiewiczACSPHOT,2014CacciolaACSNANO}
for the thin coating layer of adsorbed dyes (see Methods). This simple shell model allows us to illustrate the principle of the experimental technique, and highlight a number of difficulties to overcome in practice. We first consider in Fig.~\ref{Fig1}(a) the optical properties of a colloidal solution of 30\,nm-radius silver nanospheres in water. For a colloid concentration of 8\,pM as used in this work, the peak extinction corresponds to an optical density (OD) of 0.85\,cm$^{-1}$.
In comparison, the peak extinction of a typical chromophore has an OD of $10^{-3}$\,cm$^{-1}$ only at a concentration of 10\,nM. This concentration corresponds to sub-monolayer coverage on the colloid surface (assuming 100\% adsorption for simplicity, this translates to $\sim$1200 dyes per colloid, or 0.1\,dye/nm$^{2}$), where dye--dye interactions are expected to be negligible, and is also in the typical range for surface-enhanced spectroscopy experiments. 

A natural approach to retrieve the absorbance spectrum of the adsorbed dye would then consist in measuring the UV--Vis extinction spectrum for the mixture, and subtracting a reference taken as the bare colloids (with no adsorbed dyes). If the dye and plasmon resonance coincided, it would be difficult to disentangle them, but since the detuning between the dye and plasmon resonances is chosen to be relatively large (100\,nm), one could expect to isolate a meaningful absorbance for the dye. However, as shown in Fig.~\ref{Fig1}(b) this naive approach does not work: the predicted difference spectrum obtained from extinction measurements departs substantially from the absorbance spectrum of the adsorbed dye, showing a negative difference within the tail of the plasmon resonance, strongly distorting the lineshape around 500\,nm. This is a consequence of the minute shift of the plasmon resonance induced by the dye layer, and is comparable in magnitude with the dye absorbance itself.
This unavoidable effect is largely alleviated by considering the {\em differential absorbance} $\delta\sigma_\mathrm{abs}(\lambda)$, which results in a spectrum similar to that of the original dye. The NP absorbance is $\sim 6$ times smaller than extinction, thus allowing easier extraction of the dye contribution within the otherwise overwhelming NP scattering response. 
 A closer look at Fig.~\ref{Fig1}(b) nevertheless shows that $\delta\sigma_\mathrm{abs}(\lambda)$ is not exactly the same as that of the bare dye, in particular in the relative intensity of the main and vibronic peaks. This is attributed to the wavelength-dependence of the plasmonic enhancement, which for absorption is characterized by the surface-averaged local field intensity enhancement factor $\Mloclambda$~\cite{2009book}.
This effect is secondary and cannot cause a significant resonance shift, but can nevertheless be accounted for: the corrected differential absorbance spectrum $\delta\sigma_\mathrm{abs}(\lambda)/\Mloclambda$
then resembles very closely the true surface absorbance of the dye as seen in Fig.~\ref{Fig1}(b).

This discussion has so far focused on a low dye concentration, but in many of the previous studies dye concentrations have instead been large, typically at or beyond monolayer coverage. Figs~\ref{Fig1}(c--d) illustrate how dye--dye interaction effects may then interfere with the measurement of the intrinsic absorbance of the adsorbed dye. As the dye concentration increases the EM interaction between dyes needs to be included in the effective medium properties of the coating layer, which we describe via the Clausius-Mossotti equation (see Methods). To disentangle the effect of the NPs from the intrinsic dye--dye interactions we consider a fictitious dye-coated water nanosphere in (c) as a toy model and compare it to the same shell on a silver sphere.
In the latter case, the differential absorbance is corrected for the plasmonic enhancement as discussed earlier. Below concentrations of 0.1\,dye/nm$^3$ in the shell, local-field (Clausius-Mossotti) corrections are negligible and the differential absorbance spectra in (c-d) closely match the spectrum of the free dye in solution
(dashed red lines).
As the dye concentration increases, the dye resonance splits into two resonances as predicted from the standard exciton model~\cite{1963Kasha}: the blueshifted resonance is associated with the side-by-side dipolar interaction between dyes, while the redshifted one corresponds to head-to-tail dipoles. Both resonances are excited in a spherical shell as shown in the surface-field diagram in (c).
The same types of interactions are expected for Ag--dye core--shell structures, but the electric field is then mostly perpendicular
to the surface (see inset in (d)). 
Thus for an isotropic dye layer on a metal NP, the blue shifted (side-by-side) resonance will be favoured. 
These predictions highlight the importance of dye--dye interactions effects in dye layers.

We conclude this preliminary discussion with a summary of our findings for the determination of the intrinsic surface absorbance of dye molecules adsorbed on silver colloids. Firstly, differential absorbance measurements should be preferred over extinction. Secondly, a large detuning from the plasmon resonance simplifies the analysis, allowing direct access to the modified dye absorbance, which is only scaled by the plasmonic
enhancement factor $\Mloclambda$. Finally, low dye concentrations should be used to limit the effect of dye--dye interactions.
This combination of factors results in challenging experiments and requires pushing the limit of detection of currently available instruments with new developments.

\begin{figure*}
\centering
\includegraphics[width=14cm]{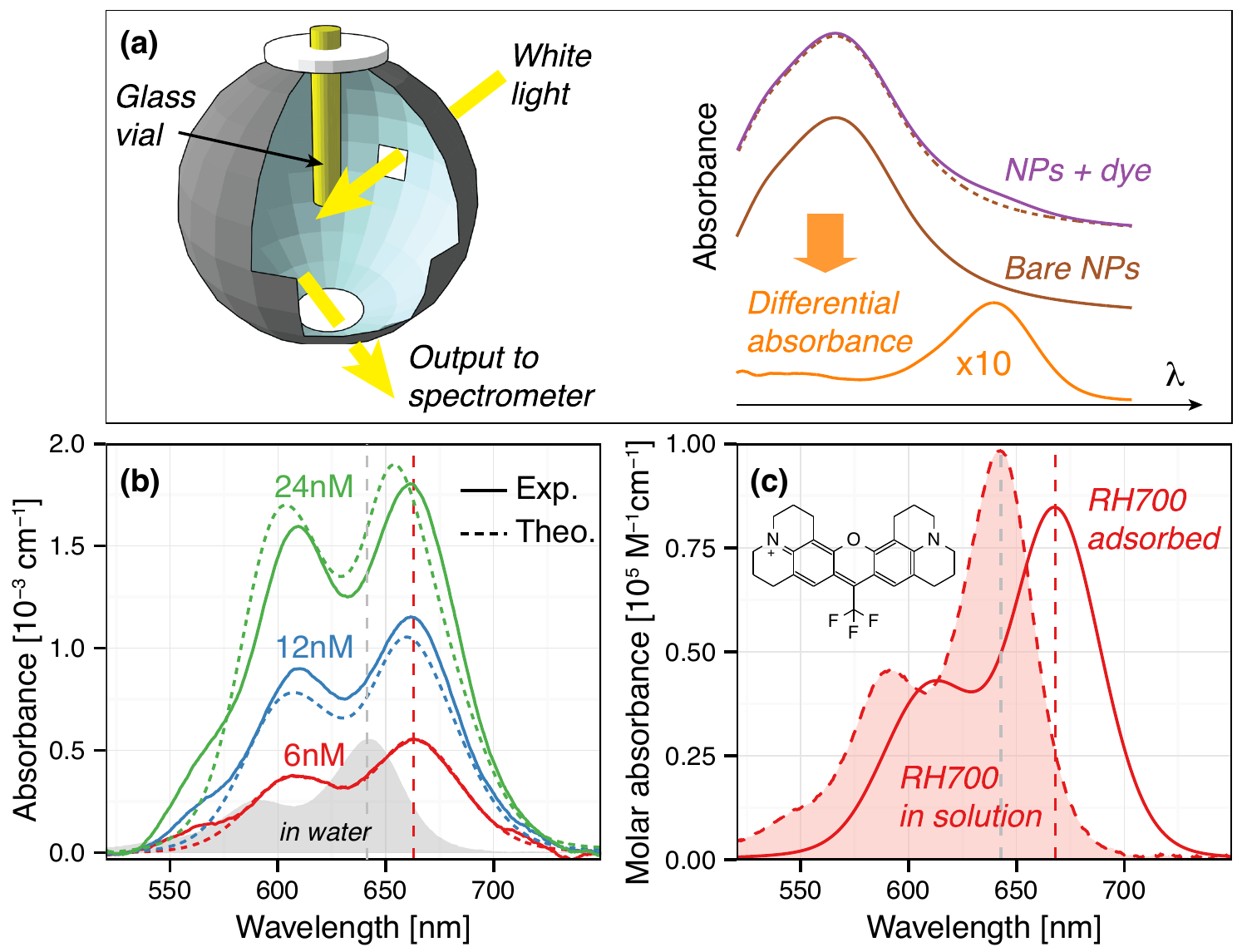}
	\caption{(a) Schematic of the set-up to measure the differential absorbance of dyes on NPs. As shown on the right, this is obtained
	by subtracting the reference absorbance spectrum of the NPs from that of an identical NP+dye solution. 
(b)	Measured (solid lines) and predicted (dashed lines) differential absorbance spectra of Rhodamine 700 (RH700) at concentrations of 6, 12, and 24\,nM on 30\,nm-radius Ag nanospheres (8\,pM).  Those spectra are clearly red-shifted by $\sim 20$\,nm compared to the normal RH700 spectrum in water (shown in grey). Note a scaling factor of $\sim 5.3$ is necessary to quantitatively match experiments and theory (see discussion in the text). The modified surface polarizability of RH700 was modelled as a double-Voigt-type resonance, with the same oscillator strength as bare RH700 and parameters adjusted to fit the experimental surface absorbance spectrum measured at 6\,nM. The effective thickness of the shell was adjusted to $L_D=0.6$\,nm. 
(c) Comparison between the deduced intrinsic surface absorption spectrum of RH700 and the reference bare RH700 (inset: chemical formula of RH700).}
\label{fig2}     
\end{figure*} 

\section*{Results}

To tackle this problem, we employ a UV--Vis absorption setup where the sample is centrally located inside an integrating sphere
(see Fig.~\ref{fig2}(a) and Methods for details). This particular configuration has only been used in rare occasions and mostly for the characterisation of seawater~\cite{1993NelsonAO,2002BabinLO,2009GaigalasJRES}.
The integrating sphere eliminates the influence of scattering, with a direct measurement of absorbance, rather than extinction. In our configuration, it
also increases the effective path-length to about 2.4\,cm (see Sec. S.II. in Supp.~Info.). With appropriate precautions (see Methods), we managed to achieve accurate referencing of the dye--NP solution absorbance spectrum against an identical NP-only solution, with $\sim 10^{-4}$ relative accuracy. 
These developments allowed us to measure the extremely small differential absorbance spectra required for this study, with a lower limit
down to an OD of about $10^{-4}$\,cm$^{-1}$.

Fig.~\ref{fig2} highlights some of the observations that are made possible by this novel approach. Fig.~\ref{fig2}(b) shows the differential absorbance spectra of the dye Rhodamine 700 adsorbed on Ag NPs at three different concentrations (6, 12 and 24\,nM). The reference absorbance spectrum of this dye in water (shown in grey) exhibits a main resonance at 642\,nm with a vibronic shoulder at 590\,nm. 
Fig.~\ref{fig2}b clearly demonstrates the ability of our approach to measure the surface absorbance of dyes on NPs at very low concentrations, down to 6\,nM here. The most dramatic observation is the clear redshift of the surface absorption spectrum by $\sim 20$\,nm, with the main resonance now at 664\,nm. As explained earlier, this cannot to be due to the interaction with the plasmon resonance (the dye and plasmon resonances are detuned by more than 200\,nm here). Moreover, this cannot be attributed either to dye--dye interactions. The effect of dye--dye interaction is in fact visible in the concentration dependence and results in a (very small) blue-shift of the resonances along with a change in relative intensity between the two peaks at 12 and 24\,nM.
These two features are in fact also predicted by the theoretical electromagnetic model shown as dashed lines. The observed red-shift at the lowest concentration must therefore be an intrinsic feature of the adsorbed dye.

The measured spectrum at 6\,nM remains affected by the plasmonic enhancement factor $\Mloclambda$, which is
easily accounted for using the effective shell model. By fitting the experimental surface absorbance spectrum measured at 6\,nM to this model,
we can therefore deduce the modified intrinsic surface polarizability of RH700 (see Sec. S.III. in Supp.~Info. for details),
as shown in Fig.~\ref{fig2}(c).
However, the oscillator strength for this modified transition appeared at first $\sim\!\! 5.3$ times smaller than that of the original spectrum. While we cannot exclude this possibility, it is more likely that the oscillator strength is unchanged and that the magnitude of the actual absorption enhancement is smaller than predicted (by the same factor $\sim\!\!5.3$).
In fact, electromagnetic theory predicts that for a 30\,nm-radius Ag sphere, the average absorption should be enhanced by a
factor of $\Mloc = 6.4$ at 664\,nm but this implicitly assumes an isotropic polarizable dipole.
Most dyes however (including RH700) have a strongly uniaxial transition dipole moment along the $\pi$-conjugated backbone of the molecule. Because the local electric field is primarily perpendicular to the metal surface (with only $\Mlocparallel=0.17$ for the parallel component at 664\,nm), the observed enhancement in absorption will be strongly dependent on the preferential orientation of the adsorbed dye. This effect is similar to surface-selection rules in surface-enhanced
spectroscopies \cite{1985MoskovitsRMP,2011LeRuChemComm}. The observed reduced absorption enhancement can therefore be attributed to an adsorption orientation preferentially flat (or at a small angle) on the surface, which is expected for dyes with such an aromatic backbone.
In this context, the assumption of a conserved oscillator strength appears the most natural and the deduced modified absorbance
is shown in Fig.~\ref{fig2}(c) along with that of RH700 in water. This ``surface absorbance'' represents the intrinsic absorption spectrum of the isolated adsorbed dye.

This measured modified absorbance can then be used to predict within the Mie--shell model the concentration dependence as shown in Fig.~\ref{fig2}b (dashed lines).
The shell thickness, $L_D$, which is an``effective'' monolayer--thickness in this model~\cite{1973DignamJCS}, was adjusted to $L_D=0.6$\,nm to reproduce the observed change in the relative intensity of the two peaks. The predicted spectra are in close agreement with the experimental results, barring the scaling factor of $5.3$ previously discussed. One may nevertheless notice that the theory predicts a larger blue shift at high concentrations than observed experimentally. This discrepancy is consistent with the expectation that the dye is preferentially adsorbed flat: as discussed in Fig.~\ref{Fig1}(d), the blueshift is related to side-by-side interactions between induced dipole perpendicular to the metal surface, and would be much less prevalent for flat adsorption. The general agreement between theory and experiment in Fig.~\ref{fig2}(b) provides strong evidence that the observed changes in relative peak intensities can be attributed to dye--dye interactions and that these effects can remain important even at very low concentrations, here down to $\sim 10$\,nM, in a regime where many surface-enhanced spectroscopy experiments are carried out.

\begin{figure*}
		\begin{center} 
				\includegraphics[width=14cm]{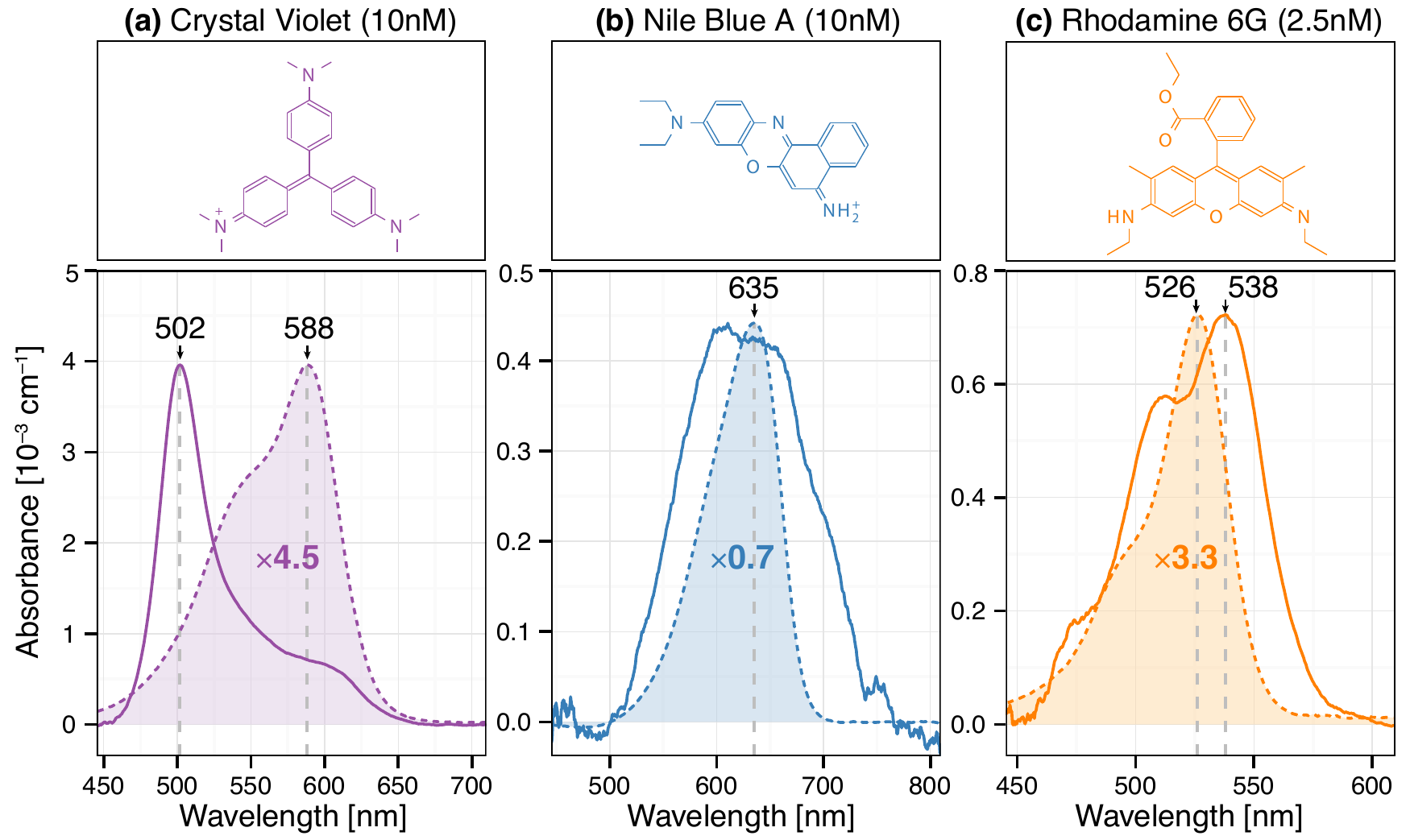}
		\end{center}
		\caption{Differential absorbance spectra of (a) Crystal Violet, (b) Nile Blue A, and (c) Rhodamine 6G adsorbed on Ag NPs (concentration 8\,pM). The dye concentrations are low enough (10, 10, and 2.5nM respectively) to avoid any effects from dye-dye interactions. The dashed lines are the reference spectra
		in water that would be measured at the same concentration, scaled to their maximum for easier visualization. The oscillator strength enhancements
		in each case are 3.4, 1.2, and 4.4, respectively. The dye chemical formula are reproduced at the top of each panel
		for convenience.}
\label{fig3}     
\end{figure*}

We have also extended this study to other chromophores commonly used in surface-enhanced spectroscopies, namely Crystal Violet (CV), Nile Blue (NB), and Rhodamine 6G (RH6G) adsorbed onto the same Ag colloids. Those complementary results are summarised in Fig.~\ref{fig3}. Of the four dyes, CV exhibits the most noticeable change in absorption, with a $\sim$ 90\,nm blue shift to 500\,nm from its main absorption peak in water at 590\,nm. Such a dramatic spectral shift suggests a major chemical change for the adsorbed species. CV is known to be sensitive to pH \cite{angeloni1979resonance}, because the number of protonated nitrogen groups strongly affects its resonance. We speculate that chemical binding to silver may have a similar effect here.
In contrast to CV, NB shows no major resonance shift, which may be interpreted as a weaker interaction with the surface through physisorption.
A broadening of the absorbance spectrum at long wavelengths is also observed and could be attributed to inhomogeneous broadening as a result of varying
surface/dye interactions from one molecule to another.
Finally, the observations for RH6G are comparable to those of RH700, except that even lower concentrations (down to 2.5\,nM) are necessary to reach the regime where dye--dye interactions become negligible. In fact, at 10\,nM, the spectrum is clearly strongly modified by these effects (see Sec. S.IV. in Supp.~Info). This may be because dimer formation is more likely as suggested in previous studies~\cite{2007ZhaoJACS}, but further work is necessary to fully elucidate this aspect.
The most likely explanation for the observed red-shift for RH700 and RH6G is that the metallic surface perturbs the electronic structure --as would any dielectric environment--, as predicted from DFT calculations~\cite{2010MortonJCP}.

In terms of absolute spectrum intensity, NB exhibits an absorption enhancement factor comparable to that of RH700, i.e of the order of 1.2 assuming conserved oscillator strength and much smaller than the predicted isotropic enhancement of $\Mloc \sim 6.4$ at 650\,nm. This is consistent
with a flat adsorption geometry, as independently confirmed for NB in air on Au surfaces~\cite{2011LeRuChemComm}.
CV and RH6G exhibit larger apparent absorption enhancements of the order of 4, the largest (assuming a preserved oscillator strength) being for 
RH6G at $\sim 4.4$. This smaller discrepancy with the predicted isotropic enhancement of $\Mloc \sim 12$ at 540\,nm indicates that RH6G does not
adsorb with its electronic transition dipole exactly parallel to the metal surface like RH700 and NB.
This is consistent with its non-planar geometry due to its
additional ring oriented perpendicular to its main backbone.
Further extensions of the EM model to anisotropic and inhomogeneous dye layers
will be necessary to confirm the importance of orientational effects.

\section*{Discussion and conclusion}
Our proposed technique, based on the measurement of the differential absorbance of a sample inside an integrating sphere, provides a unique
approach to experimentally access the surface absorbance spectra of chromophores adsorbed on NPs.
Our results clearly highlight the great potential of this technique for ultra-sensitive absorbance measurements in the presence of strongly scattering media, such as solutions of metallic colloids.
The ability to measure the intrinsic absorbance of dyes on metallic NPs paves the way for more detailed studies of
both the dye--surface and dye--dye interactions. Recent related works in this area~\cite{2012Forker,2015DietzeJPCC} have been limited to adsorbed molecules on flat metal films and exposed to air, using polarization- and/or angle-dependent transmittance/reflectance spectroscopy. In such studies, it is extremely challenging to reliably control the molecular surface coverage/concentration because of the solution-to-surface transfer step. Moreover, the ambient medium can have a strong effect on the absorbance of dyes and the results obtained in air cannot be easily compared to the reference measured in water. 
Our method mitigates both of these issues by working in solution and can reliably reach sub-monolayer concentrations.

Two important conclusions stand out from this study, which we expect to have many implications. Firstly, modifications of the intrinsic dye polarizability upon adsorption on metallic NPs appear to be the norm rather than the exception, and secondly, further modifications arising from dye--dye interactions remain important even at relatively low surface coverage, in fact in a range relevant to many experiments.
Both of these effects should be carefully considered for any experiments involving dyes adsorbed on the surface of metallic NPs, notably molecular plasmonics (strong coupling, plexcitonics) and SERS, SEF, and other surface-enhanced spectroscopies. As an example, the experimental determination of SERS enhancement factors \cite{2013LeRuMRS} typically assumes that the polarizability of the probe molecule is not modified upon adsorption.
In view of our results, this assumption should clearly be revisited for experiments in resonance or pre-resonance conditions, with
implications for the debate over the ``chemical enhancement'' mechanism~\cite{2013MoskovitsPCCP}.
Similarly, in the context of molecular plasmonics, most studies are interpreted in terms of coupling between dye and plasmon resonances,
where the dye (or $J$-aggregate) spectrum is assumed to be unchanged in contact with the metal. While this may be the case for some molecules, our results suggest that such an assumption should be carefully checked on a case-by-case basis, and that dye--dye interactions and/or intrinsic absorbance modification may play a major role in interpreting experimental results.

Beyond those direct consequences, we anticipate that our experimental findings will spark a fruitful discussion on the interpretation
of the observed spectral changes for different molecules.
New experimental access to such a regime of molecule--metal interaction with a non-obtrusive optical technique will undoubtedly entice further 
tests and refinements of DFT predictions of the resonance of dye--metal complexes. It will also offer a deeper insight into the subtle interplay between electromagnetic effects, including selection rules linked to preferential orientation on the surface,
and those of a more chemical nature attributed to changes in the electronic configuration of a molecule physisorbed or chemisorbed onto a surface. A refined understanding of these mechanisms will have a profound impact in the interpretation and further development of  surface-enhanced Raman scattering, chiroptical, or fluorescence spectroscopies.
\section*{Methods}
\noindent{\bf Materials.}
All dyes and nanoparticles were used as purchased and diluted appropriately. Rhodamine 700, Rhodamine 6G, Nile Blue, and Crystal Violet were purchased from Sigma Aldrich in powder form and mixed with ultrapure MilliQ water to the desired stock concentration (in the range 10--100\,$\mu$M). 
Stock solutions of 60\,nm diameter citrate-covered Ag nanospheres in water (particle concentration $1.9\times 10^{10}$\,mL$^{-1}$, equivalent to 32\,pM) were purchased from Nanocomposix and stored sealed in a fridge at 4$^\circ$C.
\\
\noindent{\bf Absorbance measurements.}
A modified Ocean Optics ISP-80 integrating sphere was used for all absorbance measurements. 2\,mL of solution to be measured were pipetted into a cylindrical glass tube (67.5\,mm long, 7\,mm external diameter, wall thickness 1\,mm), and were carefully inserted into the integrating sphere through a port in the north pole. A custom holder was used to ensure reproducibility of the tube position and orientation. White light illumination was provided by a 100\,W halogen lamp (See Fig. S2 for the lamp spectrum) and delivered to the sphere's equatorial port via a ThorLabs 1000\,$\mu$m 0.22 NA optical fibre. Collection was made via a similar optical fiber (910\,$\mu$m, 0.22 NA) inserted into a custom-drilled detection port in the bottom of the sphere, and coupled to an Ocean Optics USB-2000+ spectrometer with 100\,$\mu$m entrance slits and a detection range of 200\,nm to 850\,nm. Spectra were acquired as the average of $10^4$ individual 1\,ms spectra (giving an effective total exposure of 10\,s) and corrected with identically acquired dark spectra; care was taken to avoid the non-linear region of the detector, keeping counts near $\frac{1}{4}$ of the CCD pixel well-depth.
\\
Absorbance spectra are then obtained from the standard relation $A(\lambda)=-\log_{10}(I_S(\lambda)/I_R(\lambda))$, where $I_S$ and $I_R$ are the detected
intensities of the sample and the reference. For the differential absorbance of adsorbed dyes on NPs, the reference is the same NP solution with no dyes.
All differential absorbance spectra are then corrected by the effective path--length, $L_\mathrm{sphere}=2.4$\,cm to yield the optical
density in cm$^{-1}$. $L_\mathrm{sphere}$ was determined by measuring the absorbance of
Eosin B and Nile Blue in water inside the sphere (Fig.~S3) and comparing to the absorbance measured in a standard UV-Vis setup.
Moreover, since Eosin B is negatively charged, it does not adsorb to the Ag NPs and its absorbance spectrum inside the sphere
is unchanged when measured in colloidal solution (Fig.~S3).
This confirms that the strong scattering of the colloidal solution has no effect on the differential absorbance spectra.
See Sec. S.II. of Supp.~Info. for further details.\\
\noindent{\bf Sample preparation.}
All colloid samples were prepared according to the guidelines outlined in \cite{2014DarbyJACS}. Each set of measurements requires a reference solution. Reference samples were prepared by first mixing 500\,$\mu$L of Ag colloid solution with 500\,$\mu$L of 2\,mM KCl to displace the citrate capping layer (without inducing aggregation \cite{2014DarbyJACS}). This solution was then mixed with 1\,mL of MilliQ water and the resulting 2\,mL (with a final colloid concentration of 8\,pM) was pipetted into a cylindrical glass tube. The same procedure was performed for dye--colloid samples but replacing the water in the final dilution by twice the desired concentration of dye. For each concentration series, the same vial was used to ensure the exact same vial geometry for all measurements and concentrations were tested going from reference (no dye) to the highest dye concentration sequentially. Vials were cleaned with aqua regia prior to measuring and were rinsed thoroughly with water and ethanol after each sample was measured.\\
Note that this method relies on the assumption that the addition of dyes (which are charged) to colloids does not induce any intrinsic change in the colloid, in particular aggregation. Were aggregation present in the dye--colloid sample, referencing with bare colloids would no longer be valid.
For the dye concentrations and the type of colloids used here, aggregation is not expected and would be evident in the appearance of a longer wavelength absorption peak associated with NP aggregates which is not present in the spectra measured.
\\
\noindent{\bf Theory.}
Theoretical predictions of the optical properties of the dye layer were carried out using an isotropic shell model as used in most
recent studies of strong coupling in molecular plasmonics~\cite{2004WiederrechtNL,2008FofangNL,2010NiNL,2012ChenJPCC,2013ZenginSREP,2013SchlatherNL,2014AntosiewiczACSPHOT,2014CacciolaACSNANO}. This consists in solving the electromagnetic scattering problem for a metallic nanosphere
covered by a thin spherical shell (of thickness $L_D$) with an effective dielectric function that models the dye
response. Note that $L_D$ can here be viewed as an effective thickness, only loosely connected to the physical size of the dye molecule, i.e.
it is an adjustable parameter (whose meaning is defined more precisely in Ref.~\cite{1973DignamJCS}). In order to study the dye concentration dependence, it is necessary to start from the polarizability of the individual dyes
and account for local field corrections using the Clausius-Mossotti (CM) relation in the 2D dense dye layer, as one would do in a bulk 3D phase.
This has been studied in the past \cite{1973DignamJCS} and we adapted it to the case where the dyes are embedded in a medium,
here water with a dielectric constant $\epsilon_M=1.77$. The effective
dielectric function of the dye layer is then 
\begin{equation}
\epsilon_\mathrm{dye} = \frac{1+\frac{2}{3}(\tilde{\alpha}_M+\tilde{\alpha}_D)}{1-\frac{1}{3}(\tilde{\alpha}_M+\tilde{\alpha}_D)}
\end{equation}
where $\tilde{\alpha}_M$ and $\tilde{\alpha}_D$ are the normalized bare polarizabilities of the solvent and dye molecules. The former
is simply deduced from the standard CM equation in a pure phase,
\begin{equation}
\tilde{\alpha}_M=3\frac{\epsilon_M-1}{\epsilon_M+2}(\approx 0.61).
\end{equation}
The latter depends directly on the density of dyes in the shell layer, via the concentration of dyes, $c_D$ (note that this is related to the number of dyes per unit area $\mu_D$ by $c_D\approx \mu_D/L_D$):
\begin{equation}
\tilde{\alpha}_D(\omega) = c_D \frac{\alpha_D}{\epsilon_0}.
\end{equation}
The bare polarizability $\alpha_D(\omega)$ can be deduced from the measured optical absorbance of the dye up to a constant related to
the static polarizability, which is calculated by DFT (see Sec. S.I. of Supp. Info. for details).\\
The electromagnetic problem is solved using Mie theory for multilayered spheres using standard codes \cite{2009SPlaC}.
The dielectric function of silver is taken from the analytical fit in Ref. \cite{2009book}.
$\Mloclambda$ is calculated for the NP-only model, by taking the average of the field intensity enhancement factor
on the spherical surface in the middle of the thin shell.
\\[0.5cm]

\noindent{\bf \large Acknowledgements}\\
ECLR is indebted to the Royal Society of New Zealand (RSNZ)
for support through a Rutherford Discovery Fellowship. The authors are grateful
to Andy Edgar, Peter Northcote, and Matthias Lein from Victoria University of Wellington
for fruitful discussions.\\

\noindent{\bf \large Author Contributions}\\
BLD and ECLR designed the original ideas presented in this work and built the
experimental set-up with AEP. BLD carried out most of the experiments.
MM optimized various aspects of the experimental set-up, carried out extensive
calibration, and performed the DFT calculations. BA, BLD, and ECLR developed and performed
the electromagnetic theory and calculations. The manuscript was jointly written by BLD, BA, and ECLR.
All authors discussed the results and the manuscript.\\

\noindent{\bf \large Competing financial interest}\\
The authors declare no competing financial interests.\\

\noindent{\bf \large Supplementary information}\\
Supplementary information is available for this paper.

\bibliographystyle{nature2}
\bibliography{RefsAbsorbance}

\end{document}